\documentclass[preprint, superscriptaddress, floatfix, preprintnumbers, nofootinbib, altaffilletter]{revtex4-1}
\usepackage{graphicx}
\usepackage{subfigure}
\usepackage{dcolumn}
\usepackage{bm}
\usepackage{amssymb}
\usepackage{amsmath}
\usepackage{xcolor}
\usepackage[utf8]{inputenc}
\usepackage[OT1]{fontenc}
\usepackage{yfonts,amsthm,amsfonts,amssymb,amscd, ulem}
\usepackage{enumerate}
\usepackage{fancyhdr}
\usepackage{mathtools}
\usepackage{mathrsfs}
\usepackage{cancel}
\usepackage{slashed}
\usepackage{bigints}
\usepackage[flushleft]{threeparttable}
\usepackage[colorlinks=true, citecolor=purple, linkcolor=blue]{hyperref}
\usepackage{url}
\usepackage{makecell,booktabs}
\usepackage{braket}
\usepackage{relsize}
\usepackage{multirow}
\usepackage{verbatim}
\usepackage{txfonts}
\usepackage{upgreek}
\usepackage{extarrows}
\usepackage{array}
\usepackage{appendix}
\usepackage[T1]{fontenc}
\usepackage{setspace}

\begin{document}

\title{Probing Scalar-Mediated Sterile Neutrinos with Gravitational Wave and Colliders Signals}

\author{Qi Bi}
\email{biqii@buaa.edu.cn}
\affiliation{School of Physics, Beihang University, Beijing 100083, China}

\author{Jinhui Guo}
\email{guojh23@buaa.edu.cn}
\affiliation{School of Physics, Beihang University, Beijing 100083, China}

\author{Jian Liao}
\email{liaoj35@stu.pku.edu.cn}
\affiliation{School of Physics and State Key Laboratory of Nuclear Physics and Technology, Peking University, Beijing 100871, China}

\author{Jia Liu}
\email{jialiu@pku.edu.cn}
\affiliation{School of Physics and State Key Laboratory of Nuclear Physics and Technology, Peking University, Beijing 100871, China}
\affiliation{Center for High Energy Physics, Peking University, Beijing 100871, China}

\author{Xiao-Ping Wang}
\email{hcwangxiaoping@buaa.edu.cn}
\affiliation{School of Physics, Beihang University, Beijing 100083, China}

\preprint{$\begin{gathered}\includegraphics[width=0.05\textwidth]{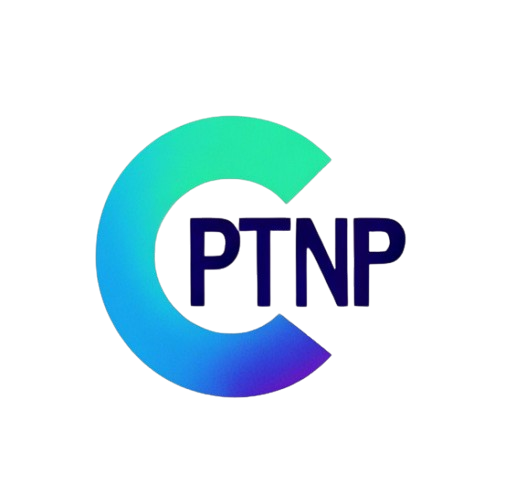}\end{gathered}$\, CPTNP-2025-037}

\begin{abstract}
We propose a UV-complete extension of the Standard Model in which a gauge-singlet scalar $S$ acquires a vacuum expectation value, generates a Majorana mass for a sterile neutrino $N$, and mixes with the Higgs field. This framework addresses neutrino masses via a seesaw mechanism and, for sufficiently large scalar mixing, can also drive a strong first-order electroweak phase transition, producing gravitational-wave (GW) signals potentially detectable by GW observatories. The Higgs–$S$ mixing also enhances sterile-neutrino pair production at colliders through $s$-channel exchange of the Higgs and $S$. Owing to the small active–sterile mixing angle, $N$ is generically long-lived, yielding characteristic displaced-vertex signatures. The combination of GW observations and displaced-vertex searches at colliders provides complementary cross-checks of the model parameter space.
\end{abstract}

\maketitle


\section{Introduction}
\label{sec:int}

As one of the most straightforward extensions to the standard model (SM), the sterile neutrino, also known as heavy right-handed neutrinos, obtains tremendous attention and investigation due to its natural solution to the neutrino mass puzzle via the well-known seesaw mechanism \cite{Schechter:1980gr,Ma:2006km,Foot:1988aq,King:2013eh,Mohapatra:1979ia,Gell-Mann:1979vob}. The mechanism is named after a seesaw because the masses of the light left-handed neutrinos and the sterile neutrino are inversely related. Building on this basic idea, several variants of the seesaw mechanisms are developed, such as the type-I, type-II, hybrid, inverse, and linear seesaw mechanisms \cite{Schechter:1981cv,Akhmedov:1995ip,Akhmedov:1995vm,Barr:2003nn}. For the type-I seesaw, the mass $m_{\nu_s}$ of the sterile neutrino and the mixing angle $\theta$ between the active and the sterile neutrino are fixed by $\theta\sim\sqrt{m_{\nu_{\rm SM}}/m_{\nu_s}}$, where $m_{\nu_{\rm SM}}$ is the SM neutrino mass. Given the upper limits on the active neutrino mass, $m_{\nu_{\rm SM}}<0.1$ eV, the mixing angle is constrained to be approximately $\theta\sim \sqrt{m_{\nu_{\rm SM}}/m_{\nu_s}}<10^{-6}\cdot \sqrt{\frac{100{\rm GeV}}{m_{\nu_s}}}$ \cite{Planck:2018vyg,DiValentino:2021hoh}. This extremely tiny mixing angle makes the study of sterile neutrinos at colliders challenging, especially when their masses are heavier than those of the $W$ and $Z$ bosons. 

In recent years, numerous searches have been implemented to constrain the possible mass range and interaction form of sterile neutrinos via collider, cosmology, dedicated neutrino measurement and other relevant experiments \cite{Kopp:2013vaa,MicroBooNE:2015bmn,Dentler:2018sju,NEOS:2016wee,IceCube:2016rnb,Dasgupta:2013zpn,Abazajian:2017tcc,Super-Kamiokande:2014ndf,Smithers:2021orb,Arguelles:2021gwv,Goldhagen:2021kxe,Dekens:2021qch,Camara:2020efq,Seo:2020fdu}, which greatly enriched and deepened people's understanding of sterile neutrinos. Among these experiments, high-energy colliders, the most feasible and reliable platform to explore sterile neutrinos, are a very active research area in particle physics. Generally, in terms of the experimental level, in order to probe the sterile neutrinos, their mixing angles with active neutrinos and their masses are usually taken to be independent, as a result, plenty of model-independent searches are applied in a broad mass range from eV to TeV. For the case below GeV, peak searches in the energy spectra of charged pseudoscalar-meson leptonic decays \cite{Britton:1992pg}, lepton universality ratio tests \cite{Bryman:2019ssi,Bryman:2019bjg}, and lepton number violation investigations \cite{CLEO:2010ksb,BELLE:2011bej,LHCb:2014osd,BESIII:2019oef} have been explored. For masses above GeV and lighter than SM gauge boson, many studies are focusing on its long-lived signature searches at CMS and ATLAS \cite{CMS:2022fut,ATLAS:2022atq,Bondarenko:2019tss,CMS:2024hik,Helo:2013esa,Liu:2019ayx,ATLAS:2025uah,ATLAS:2024ocv,Urquia-Calderon:2023dkf,Cottin:2022nwp}, where its long-lived feature usually comes from the three-body decays. Their direct production is usually difficult for masses above the Higgs boson due to the heavy mass and tiny mixing angle. In Ref. \cite{Yang:2023ice}, the authors discussed the potential to directly probe this heavy sterile neutrino at the future electron-proton (Ion) colliders, such as LHeC and EIC. Moreover, over the past few years, due to the relatively small SM background and high effective center-of-mass energy of muon colliders \cite{Neuffer:1986dg,Ankenbrandt:1999cta,Boscolo:2018ytm,Long:2020wfp,Black:2022cth,Accettura:2023ked}, some projected predictions at muon colliders have been made regarding the possibility of sterile neutrinos \cite{Li:2022kkc,Chakraborty:2022pcc,Mekala:2023diu,Kwok:2023dck,Li:2023tbx,Li:2023lkl,He:2024dwh,Bi:2024pkk}.

Apart from the collider searches, people also have discussed the possible generating mechanism of sterile neutrino as the dark matter from first-order phase transition (FOPT) in the early universe \cite{Shaposhnikov:2023hrx,Ma:2024ymi}, where the specific generation mechanism from Q-ball decays formed from FOPT, or from the resonance with the help of lepton asymmetry, or from the first-order QCD phase transitions are investigated. Since sterile neutrinos are required to be dark matter, their masses are typically light to ensure their stability. Besides, they also discussed the possible gravitational wave (GW) detection in future GW observations, which was formed during the phase transition. These studies offer new insights into the generation mechanism of sterile neutrino dark matter and provide potential exploration methods utilizing the present and future gravitational wave observations. Consequently, it is reasonable to drill into the possible research of sterile neutrinos not only at the ground-based operable colliders but also in the gravitational wave experiments once there is an additional scalar potential\cite{Chung:2010cd,Bian:2018mkl,DiBari:2020bvn,Borah:2023zsb}.

In terms of the typical type-I seesaw mechanism, a Majorana mass term can naturally arise after the spontaneous symmetry breaking of a new scalar, so it comes very naturally to consider its own potential and mix with the potential of the SM Higgs field. Therefore, a possible electroweak FOPT of the Higgs and new scalar field may occur in the early universe, which could potentially leave detectable GW signatures in the present universe. In addition to providing Majorana masses for sterile neutrinos, introducing a new scalar field also offers a well-promising channel for producing heavy sterile neutrinos in colliders through mixing with the Higgs field. As a result, we can cross-check this mechanism at colliders and GW observations. 

Contrary to previous studies \cite{liu_testing_2022, liu_shedding_2022, liu_searching_2021, xie_lepton-mediated_2021, deppisch_heavy_2019, das_heavy_2022, carena_electroweak_2020, jana_displaced_2018}, in this work, we try to investigate a UV-complete model featuring a sterile neutrino $N$ obtaining its Majorana mass after the spontaneous symmetry breaking of a new scalar $S$, where the new scalar shares a mixing potential with the SM Higgs. The symmetry of the mixing potential between the new scalar field and the Higgs field will spontaneously break at the electroweak energy scale as the universe expands, which may result in a first-order phase transition. Therefore, it is possible to probe the associated GWs from the FOPT at experiments, such as LISA, Taiji, and TianQin. In addition, the sterile neutrino will obtain a huge Majorana mass after symmetry breaking, which can, just like the traditional type-I seesaw mechanism, generate a small mass for active neutrinos. The sterile neutrino will be naturally long-lived due to the suppression of the tiny mixing angle. And the mixing between the scalar $S$ and the Higgs field provides a large production cross section for the sterile neutrino via the Higgs and the new scalar. Firstly, we will calculate and simulate the phase transition process, get its corresponding gravitational wave spectrum, and then examine its possible detection on the GW platforms. Secondly, we explore the potential to probe the long-lived sterile neutrino at the future HL-LHC with center-of-energy $\sqrt{s}=14$ TeV and luminosity $\mathcal{L}=3~{\rm ab^{-1}}$ \cite{ZurbanoFernandez:2020cco,Solovyanov:2024gri,FernandezPerezTomei:2024xid}, concentrating on its pair production via the $s$-channel processes mediated by Higgs or new $S$ particles. Then its subsequent long-lived decay processes, such as $N\to Z\nu$, $N\to \ell W$, and $N\to h\nu$, are investigated and an inclusive search for displaced vertex signatures is implemented at the future CMS detector employing the high-resolution tracker and upgraded timing layers systems. These cross-checked searches across two different time scales and two different detection methods can cover most of the parameter space of the model, enhance the understanding of sterile neutrinos, and fill gaps in previous research.

This paper is organized as follows. In Sec. \ref{sec:model}, the specific model of the sterile neutrino in the mixed new scalar and Higgs field potential, as well as their relevant decay processes are described and calculated.
In Sec. \ref{sec:constraints}, we investigate the possible constraints on the new scalar and sterile neutrino from the collider searches and muon $g-2$.
In Sec. \ref{sec:FOPT}, the strong first-order phase transition in the early universe, as well as the gravitational waves generated by this phase transition, are then discussed.
In Sec. \ref{sec:sensi}, we discuss the long-lived signatures of sterile neutrinos and their projected sensitivities at the HL-LHC.
In Sec. \ref{sec:interplay}, we discuss the interplay between GW detections and collider searches.
In Sec. \ref{sec:conclusion}, we conclude.

\section{The model}
\label{sec:model}
To generate a mass for SM neutrinos, we extend the SM by introducing two new fields: a Majorana fermion $\hat N$ and a real scalar $\hat S$, both of which are singlets under the SM gauge group. The relevant Lagrangian is
\begin{equation}\label{eq:lagrangian}
\begin{aligned}
\mathcal{L}_N &= \bar{\hat N}i\slashed{\partial} \hat N - \left(y' \bar{\hat L}\tilde{\hat H} \hat N + \frac{y_N}{2} \hat S \bar{\hat N}^c \hat N + {\rm h.c.}\right) + \frac{1}{2}\partial_\mu \hat S\partial^\mu \hat S - V(\hat H,\hat S), \\
V(\hat H,\hat S) &= -\mu_H^2 |\hat H|^2 + \lambda_H |\hat H|^4 + \mu_1^3 \hat S + \frac{\mu_2^2}{2} \hat S^2 + \frac{\mu_3}{3} \hat S^3 + \frac{\lambda_S}{4} \hat S^4 + \frac{\mu_{HS}}{2}|\hat H|^2 \hat S + \frac{\lambda_{HS}}{2}|\hat H|^2 \hat S^2,
\end{aligned}    
\end{equation}
where $\hat H$ is the SM Higgs doublet, $\hat L$ represents the SM left-handed lepton doublet, and $V(\hat H,\hat S)$ is the potential of Higgs and the new scalar field. For concreteness, and without loss of generality, we focus on the second-generation lepton. In the following, $\nu$ ($\ell$) refers to $\nu_\mu$ ($\mu$) unless stated otherwise. After electroweak symmetry breaking, both the Higgs doublet $\hat H$ and the new scalar $\hat S$ acquire vacuum expectation values (vevs),
\begin{align}
    \hat H = \frac{1}{\sqrt{2}}\left( \begin{array}{c}
         0 \\
         v+ \hat h 
    \end{array}\right), ~~\hat S = v_s + \hat s,
\end{align}
where $v$ and $v_s$ denote the vevs of the $\hat H$ and $\hat S$ field, respectively, with $v=246 \rm ~GeV$ being the SM Higgs vev. Here, $\hat h$ denotes the neutral Higgs excitation and $\hat s$ denotes the scalar excitation of $\hat S$, both in the interaction basis. In the following subsections, we first diagonalize the scalar and neutrino mass matrices to obtain the physical states, and then analyze the relevant decay channels of the new particles.
\subsection{Scalar Part}
After spontaneous symmetry breaking, the interaction terms $|\hat H|^2 \hat S$ and $|\hat H|^2 \hat S^2$ in the scalar potential generate mass mixing between the Higgs field $\hat h$ and the new scalar $\hat s$. The mass term in the Lagrangian can be written as
\begin{equation}
    \mathcal{L}_{\rm mass}^{\rm scalar}=\frac{1}{2}\left(\hat h, \hat s \right)\left(\begin{array}{cc}
        m_H^2 & m_{HS}^2 \\
        m_{HS}^2 &  m_S^2
    \end{array}\right)\binom{\hat h}{\hat s}=\frac{1}{2}\left(\hat h, \hat s \right)M_{HS}^2\binom{\hat h}{\hat s}=\frac{1}{2}\left(h, s \right)U^\dagger M_{HS}^2 U \binom{h}{s},   
\end{equation}
where $h$ is the SM Higgs boson with $m_{h} = 125 ~\rm GeV$, $s$ is the new scalar. The explicit form of the mass matrix elements can be given by
\begin{equation}
\begin{aligned}
    m_H^2 &\equiv \frac{\partial^2 V}{\partial \hat H^2} = 2\lambda_H v^2, \\
    m_{HS}^2 &\equiv \frac{\partial^2 V}{\partial \hat S^2} = (\mu_{HS} + 2\lambda_{HS} v_s)\frac{v}{2}, \\
    m_S^2 &\equiv \frac{\partial^2 V}{\partial \hat H \partial \hat S} = -\frac{\mu_1^3}{v_s}+\mu_3 v_s + 2 \lambda_S v_s^2 - \frac{\mu_{HS}v^2}{4v_s},
\end{aligned}
\end{equation}
and $U$ is the unitary matrix that diagonalizes $M_{HS}^2$:
\begin{equation}
U^\dagger M_{HS}^2 U = \mathrm{diag}(m_h^2, m_s^2), ~ \text{with}~~ U = \left(\begin{array}{cc}
        \cos\theta_{hs} & -\sin\theta_{hs} \\
        \sin\theta_{hs} &  \cos\theta_{hs}
    \end{array}\right),
\end{equation}
The mixing angle, $\theta_{hs}$, between $h$ and $s$ can be determined by
\begin{align}
    \tan 2\theta_{hs} = \frac{2m_{HS}^2}{m_{H}^2-m_{S}^2}.
\end{align}
After diagonalizing the scalar mass matrix, the interactions of the SM Higgs and the new scalar with SM particles are modified by the mixing angle $\theta_{hs}$, such that
\begin{align}\label{eq:s-h-coupling}
    g_{h XX} = \cos\theta_{hs} \cdot g^{\rm SM}_{hXX},
    \quad  g_{s XX} = \sin\theta_{hs} \cdot g^{\rm SM}_{hXX},
\end{align}
where $g_{hXX}^{\rm SM}$ denotes the SM Higgs couplings to SM particles $X$ (leptons, quarks, and gauge bosons). Consequently, the mixing angle $\theta_{hs}$ can be constrained by precision measurements of Higgs production and decay. With $m_h = 125~\rm GeV$ and $v = 246~\rm GeV$ fixed, the scalar sector has five remaining free parameters:
\begin{align}
    \{m_{s}, \theta_{hs}, \mu_3, \lambda_{HS},  \lambda_S\}
\quad \text{or} \quad
    \{m_s, \theta_{hs}, v_s, \lambda_{HS}, \lambda_S\}.
\end{align}
Thus, we have the effective Lagrangian as 
\begin{align}
\mathcal {L}= \frac{1}{2}\partial_\mu h \partial^\mu h + \frac{1}{2}\partial_\mu s \partial^\mu s -\frac{1}{2}m_h^2 h^2 -\frac{1}{2}m_s^2 s^2 - \lambda_{shh} s h h - V_{else}(s,h),
\end{align}
where 
\begin{align}
 \lambda_{s h h} =& (\frac{1}{2}\mu_{HS} + \lambda_{HS} v_s)\cos^3\theta_{hs} + (2\lambda_{HS}v-6\lambda_Hv)\sin\theta_{hs} \cos^2\theta_{hs} \nonumber  \\
 &+ (6\lambda_S v_s + 2 \mu_3- 2\lambda_{HS}v_s -\mu_{HS})\sin^2\theta_{hs} \cos\theta_{hs}- \lambda_{HS}v\sin^3\theta_{hs}.
\end{align}
In the mass eigenbasis, the SM-like Higgs $h$ has its production cross section and decay width modified by the mixing angle $\theta_{hs}$: 
\begin{equation}
\begin{aligned}
    \sigma_{h} = \cos^2\theta_{hs} \cdot \sigma_{h}^{\rm SM}, \\
    \Gamma_{h} = \cos^2\theta_{hs} \cdot \Gamma_{h}^{\rm SM},
\end{aligned}
\end{equation}
where $\sigma_h^{\rm SM}$ and $\Gamma_h^{\rm SM}$ denote the SM Higgs production cross section and decay width. For the new scalar $s$, we consider the mass region where $m_s>2 m_h$. In this regime, $s$ can always decay into a pair of SM Higgs bosons. Consequently, we can ignore the three-body decay; the production cross section and decay width are given by
\begin{equation}
\begin{aligned}
    \sigma_s &= \sin^2\theta_{hs} \cdot \sigma_{h}^{\rm SM}(m_s), \\
    \Gamma_s &= \sin^2\theta_{hs} \cdot \Gamma_{h}^{\rm SM}(m_s) + \Gamma_{s \rightarrow h h} + \Gamma_{s \rightarrow N N},
\end{aligned}
\end{equation}
where $\sigma_h^{\rm SM}(m_s)$ and $\Gamma_h^{\rm SM}(m_s)$ represent the SM Higgs production cross section and decay width evaluated at the mass of $s$. The additional decay channels from the mixing, $s\to hh$ and $s\to NN$, contribute to the total width, which can be derived as
\begin{equation}
\begin{aligned}
    &\Gamma_{s \rightarrow h h} = \frac{\lambda_{s h h}}{32\pi m_{s}}\sqrt{1-\frac{4m^2_h} {m_s^2}}, \\
    &\Gamma_{s \rightarrow NN} = \frac{m_{s}}{16\pi}\frac{m_N^2}{v_s^2}\left(1-\frac{4m^2_N} {m_s^2}\right)^{3/2}\cos^2\theta_{hs}.
\end{aligned}
\end{equation}

\subsection{Neutrino Part}
The mass terms for the neutrino after the symmetry breaking can be written as
\begin{align}
    &\mathcal{L}_{\rm mass}^N = -\frac{1}{2}\bar{\hat n}^{c} M_N \hat n + {\rm h.c.} = -\frac{1}{2}\bar{n^c} U'^\dagger M_N U' n + {\rm h.c.},
\end{align}
where $\hat n = \binom{\hat\nu}{\hat N^c}$ represents the interaction eigenstate, $n = \binom{\nu}{N^c} $ denotes the mass eigenstate. The mass matrix of active and sterile neutrinos is 
\begin{align}
    M_N = \left(\begin{array}{cc}
        0 &  m_D^0 \\
         m_D^0 &  m_N^0
    \end{array}\right),
\end{align}
with 
\begin{align}
     m_D^0 = \frac{y'v}{\sqrt{2}}, \qquad m_N^0 = y_N v_s.  
\end{align}
The mass matrix is diagonalized by the unitary matrix $U'$, according to $U'^\dagger M_N U'={\rm diag}(m_\nu, m_N)$. $U'$ can be parameterized as a simple rotation:
\begin{align}
U'=\left(\begin{array}{cc}
        \cos\theta_{\nu N} & -\sin\theta_{\nu N} \\
        \sin\theta_{\nu N} &  \cos\theta_{\nu N}
    \end{array}\right).
\end{align}
In the seesaw limit $m_N\gg m_D$, to leading order in $m_D/m_N$, the mass eigenvalues and mixing angle are approximately
\begin{align}
    m_\nu \simeq \frac{ (m_D^0)^2}{m^0_N} \simeq m^0_N \theta_{\nu N}^2,\quad m_N \simeq m^0_N+\frac{(m_D^0)^2}{m^0_N}\simeq m^0_N,\quad \theta_{\nu N} \simeq \frac{m_D^0}{m^0_N},
\end{align}
with $\theta_{\nu N}$ representing a small mixing angle between active and sterile neutrinos. Thus, the mass of the heavy sterile neutrino is approximately equal to $ m^0_N$.
The free parameters for neutrino mass and interaction are 
\begin{equation}
    \{m_N, \theta_{\nu N}\}.
\end{equation}
The measured upper bound on the active neutrino mass ($m_\nu<0.45$ eV \cite{Katrin:2024tvg}) puts a stringent constraint on the mixing angle for a given sterile neutrino mass, which reads as

\begin{equation}
\begin{aligned}
    \theta_{\nu N} =\sqrt{\frac{m_\nu}{m_N}} \lesssim \sqrt{\frac{0.1~{\rm eV}}{100~{\rm GeV}}}\cdot \sqrt{\frac{100~{\rm GeV}}{m_N}}=10^{-6}\sqrt{\frac{100~{\rm GeV}}{m_N}},
\end{aligned}    
\end{equation}
where we conservatively take $m_\nu<0.1~{\rm eV}$. For instance, if $m_N = 100$ GeV, we have $\theta_{\nu N} \lesssim 10^{-6}$. From a naturalness perspective, the Dirac mass term of the neutrino is
\begin{align}
m_D^0=\sqrt{m_\nu m_N}=10^{-4}\sqrt{\frac{m_\nu}{0.1~{\rm eV}}\cdot\frac{m_N}{100~{\rm GeV}} }~ \rm GeV.
\end{align} 
Therefore, in the following discussions, we focus on the region with $m_N > 100~\mathrm{GeV}$ and $\theta_{\nu N} < 10^{-6}$. After performing the rotation and specifying the free parameters, to the leading order, the effective interaction Lagrangian for the sterile neutrino can be expressed as
\begin{align}
    \mathcal{L}_{\rm eff}^{\rm Int} & \simeq  \frac{g_W}{\sqrt{2}}\theta_{\nu N}\cdot W_\mu\left(\bar{N^c}\gamma^\mu \ell_L +\mathrm{h.c.}\right)
      + \frac{g_W}{2\cos{\theta_W}}\theta_{\nu N}\cdot Z_\mu\left(\bar{\nu}_L \gamma^\mu N^c
    + \mathrm{h.c.}\right) \nonumber \\
   & + \frac{m_N \theta_{\nu N} }{v} (h\cos\theta_{hs}-s\sin\theta_{hs}) \left( \bar{\nu}_L N +\theta_{\nu N}\bar{\nu}_L \nu_L^c - \theta_{\nu N} \bar{N^c}N  + \rm{h.c.} \right)  \\
   &+ \frac{m_N}{v_s}(s \cos\theta_{hs} + h \sin\theta_{hs}) (\bar{N^c}N +\theta_{\nu N} \bar{N^c}\nu_L^c + \theta_{\nu N} \bar{\nu}_L N + \rm{h.c.}).
   \label{eq:fermion-int}
\end{align}

\begin{figure}[ht]
    \centering
    \includegraphics[width=0.435\textwidth]{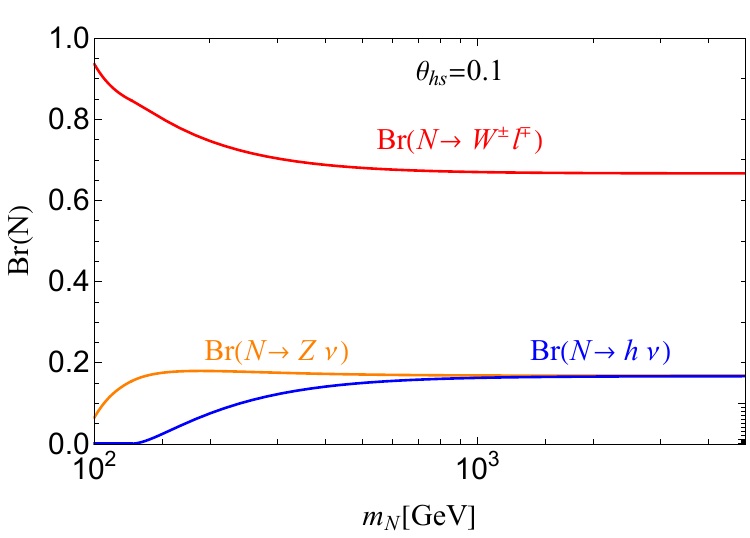}
    \includegraphics[width=0.45\textwidth]{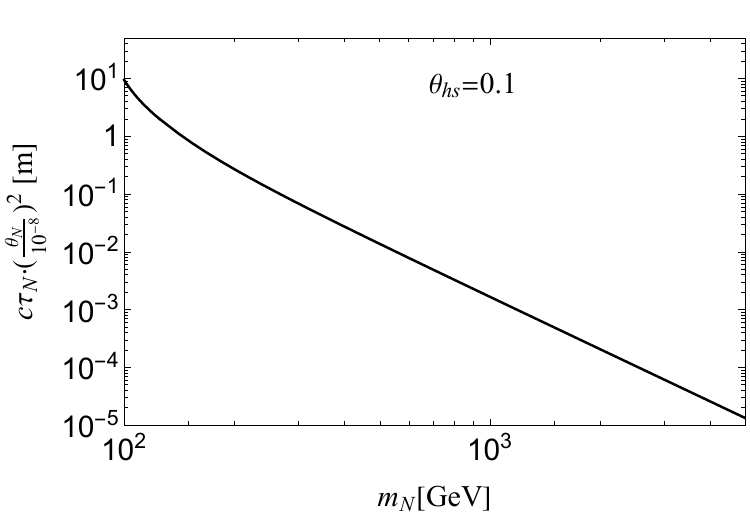}
    \caption{The branching ratios of sterile neutrino $N$ decay (left) and its proper decay length (right) as a function of sterile neutrino mass $m_N$ with $\theta_{hs}=0.1$. The three main decay branching ratios are shown in different colors with red corresponding to $N\to \ell^\pm W^\mp$, orange corresponding to $N\to \nu Z$, and blue corresponding to $N\to \nu h$.}
    \label{fig:br_NL}
\end{figure}

For our interested sterile neutrinos mass region $m_N \gg 100$ GeV and $m_s\gtrsim m_N$, the sterile neutrino can decay to $\ell W$, $\nu Z$, and $\nu h$. Thus, the total decay width of $N$ can be written as: 
\begin{align}
    \Gamma_{N} = 2 \Gamma(N \to \ell^- W^+) + \Gamma(N \to \nu_\ell Z)+\Gamma(N \to \nu_\ell h),
    \label{eq:decay-width}
\end{align}
where
\begin{equation}
\begin{aligned}
    &\Gamma(N \to \ell^- W^+)=\Gamma(N \to \ell^+ W^-)  = \frac{\theta_{\nu N}^2 g_W^2}{64 \pi} \frac{(m_N^2-m_W^2)^2(m_N^2+2m_W^2)}{m_N^3 m_W^2}, \\
    &\Gamma(N \to \nu_\ell Z) = \frac{\theta_{\nu N}^2 g_W^2}{128 \pi} \frac{(m_N^2-m_Z^2)^2(m_N^2+2m_Z^2)}{m_N^3 m_W^2}, \\
    &\Gamma(N \to \nu_\ell h) \simeq  \frac{(2v\sin\theta_{hs}+v_s\cos\theta_{hs})^2\theta_{\nu N}^2 g_W^2}{128\pi v_s^2}
    \frac{(m_N^2-m_h^2)^2}{m_N m_W^2}.
\end{aligned}
\end{equation}
In the limit $m_N\gg m_W,m_Z,m_h$ and $\theta_{hs}\ll 1$, the decay width of $N$ is proportional to $\theta_{N}^2 m_N^3 / v^2$. The branching ratios of $N$ are illustrated in the left panel of Fig.~\ref{fig:br_NL}. The label $\ell^\pm W^\mp$ denotes the sum over charge-conjugated channels, $N\to \ell^+ W^-$ and $N\to \ell^- W^+$. When $m_N$ is considerably larger than the Higgs mass, the branching ratios of $N$ are approximately: $N \to \ell^{ \pm} W^{\mp}$ at 66\%, $N \to \nu h$ at 17\%, and $N \to \nu Z$ at 17\%. This illustrates that the width of $N$ is constrained by the tiny neutrino sterile neutrino mixing angle.

When $m_N \gg m_W, m_Z, m_h$, and $\theta_{hs}\ll 1$, all decay channels of $N$ are open, the sterile neutrino's proper decay length $c \tau_N$ can be directly obtained from the total decay width of $N$ :
\begin{align}
    c\tau_N  = \frac{1}{\Gamma_{N}} \simeq 1.3 ~{\rm cm} \times \left( \frac{10^{-8}}{\theta_{\nu N}} \right)^2 \left( \frac{500 ~{\rm GeV}}{m_N} \right)^3,
\end{align}
where we fixed $g_W=0.65, m_W=80.4~ \mathrm{GeV}$, and $c$ is the speed of light. The dependence of the lifetime on $m_N$ is illustrated in the right panel of Fig. \ref{fig:br_NL}, using the complete formula of $\Gamma_N$ from Eq. (\ref{eq:decay-width}). In the relevant parameter space, for example, with $\theta_{\nu N}=10^{-8}$ and $m_N=100~ \mathrm{GeV}$, the sterile neutrino $N$ has a proper decay length of approximately 9.4 meters because the decay channel $N \rightarrow h \nu$ is forbidden and must be excluded from the calculation. This long-lived signature could potentially be observed at future HL-LHC.

In summary, we have systematically analyzed the model, focusing on the mass mixing of scalars and fermions. After diagonalizing the mass matrices, we derived the corresponding interaction Lagrangians and studied the decay widths and branching ratios of the heavy scalar $s$ and sterile neutrino $N$. The independent parameters of the model are
\begin{align}\label{eq:param}
    \{m_s, \theta_{hs}, v_s, \lambda_{HS}, \lambda_S, m_N, \theta_{\nu N}\}.
\end{align}

\section{Existing constraints}
\label{sec:constraints}
Introducing the new scalar $S$ and sterile neutrino $N$ has brought about a series of constraints in perturbation theory and collider searches. Due to the heavy mass of the sterile neutrino and its tiny mixing angle with the active neutrino, the sterile neutrino-related parts are subject to almost no constraints. The limitations arising from the interactions between the Higgs boson and the new scalar are the most significant ones. Therefore, we will mainly explore the new scalar-related constraints in this section.

From a theoretical perspective, tree-level perturbative unitarity \cite{Lee:1977eg,Luscher:1988gc} imposes a constraint on the mass and vev of the new scalar through a relation involving the partial wave amplitudes $a_J(s)$ for all possible $2\rightarrow 2$ scattering processes, which requires
\begin{align*}
    |{\rm Re}(a_J(s))| \leq \frac{1}{2},
\end{align*}
with the most stringent constraint coming from the $a_0$ partial wave. In the limit of a heavy new scalar, perturbative unitarity places a straightforward and robust upper limit on $\tan \left( \frac{v}{v_s} \right)$ due to the tiny $s-h$ mixing angle, which reads as \cite{Pruna:2013bma}
\begin{align*}
    \tan^2 \left( \frac{v}{v_s} \right) < \frac{16\pi v}{3m^2_s}, \quad {\rm for} ~a_{0}(s s \rightarrow s s) < 0.5. 
\end{align*}
For $v_s = 500~ \rm GeV$, this implies that the new scalar mass $m_s$ must be smaller than $1878~ \rm GeV$. Furthermore, perturbativity typically demands that the couplings, including $\lambda_S$, $\mu_{HS}$, and $\lambda_{HS}$, remain below $4\pi$. 

In our model, the couplings of $h$ to SM particles are reduced by the mixing angle $\theta_{hs}$, which modifies the Higgs boson production cross-sections while leaving its decay branching ratios unaffected, provided that $m_s>2m_h$.
Consequently, such parameters are stringently restricted by experimental measurements of SM Higgs signal strengths $\mu$, 
\begin{align*}
    \mu = (1 - \sin^2\theta_{hs})\cdot\mu_{\rm SM}.
\end{align*}
The joint global analysis of Higgs couplings conducted by ATLAS and CMS provides the following constraint \cite{Ellis:2013lra,Falkowski:2013dza}: 
\begin{align*}
    \mu/\mu_{\rm SM} > 0.9^2,
\end{align*}
at 95\% C.L. to account for the observed Higgs signal strengths, which requires $|\sin\theta_{hs}| < 0.43$.

Further constraints come from electroweak precision observables, particularly the high-precision measurement of the $W$ boson mass and oblique electroweak parameters $S$, $T$, and $U$.
The mass of $W$ boson, $m_W$ must align within $2\sigma$ of the experimental value $m^{\rm exp}_W = 80.385\pm 0.015 ~\rm GeV$. One-loop corrections are dependent on $m_s$ and $\theta_{hs}$. In the region of interest where the $s$ mass is heavy, the constraint on the $W$ boson mass is significantly more stringent than those obtained from the oblique parameters. These can be easily satisfied by requiring $\sin~\theta_{hs}\lesssim 0.1$ \cite{Robens:2015gla}.

\section{First-order electroweak phase transition and the gravitational waves}
\label{sec:FOPT}
In this section, we will focus on the entire electroweak phase transition process and its associated GWs. We will then explore the potential for detecting GWs in future GW observations. Specifically, the viable parameters that can result in a strong FOPT will be carefully investigated, closely related to the detection sensitivity at future colliders.

The scalar potential $V(\hat H, \hat S)$ in Eq. (\ref{eq:lagrangian}) refers to the zero temperature potential, which will acquire the thermal correction of finite temperature $T$ in the early universe. Up to $T^2$, the gauge-invariant thermally corrected effective potential takes the form \cite{Patel:2011th,Fairbairn:2013uta,Liu:2021jyc,Bian:2019bsn}:
\begin{equation}\label{eq:corr-potent}
\begin{aligned}
    V_T &= (-\mu_H^2+ c_H T^2) |\hat H|^2 + \lambda_H |\hat H|^4\\
    &+ (\mu_1^3 + m_1 T^2)\hat S + \frac{\mu_2^2 + c_S T^2}{2} \hat S^2 + \frac{\mu_3}{3} \hat S^3 + \frac{\lambda_S}{4} \hat S^4\\
    &+ \frac{\mu_{HS}}{2}|\hat H|^2 \hat S + \frac{\lambda_{HS}}{2}|\hat H|^2 \hat S^2,
\end{aligned}
\end{equation}
with 
\begin{align}\label{eq:coupling-Tem}
    c_H &= \frac{3g^2+g'^2}{16} + \frac{y_t^2}{4} + \frac{\lambda_H}{2} + \frac{\lambda_{HS}}{24},  \nonumber \\
    c_S &= \frac{\lambda_{HS}}{6} + \frac{\lambda_S}{4}+\frac{y_N^2}{12}, \nonumber \\
    m_1 &= \frac{\mu_{HS}+\mu_3}{12},
\end{align}
where $g$ and $g'$ represent the electroweak gauge couplings, and $y_t$ denotes the top Yukawa coupling. Contributions from other quarks or leptons are considered negligible. As discussed in Ref. \cite{Harigaya:2022ptp}, it is generally a good approximation to neglect small terms, such as the $y_N^2$ term in $c_S$, since it is much smaller than $\lambda_{HS}$ and $\lambda_S$. As highlighted in Ref. \cite{Liu:2021jyc}, the presence of a nonzero $\mu_1^3$, known as the tadpole term, has a significant impact on the first-order electroweak phase transition (FOEWPT) pattern. Consequently, this tadpole term is generally retained.

The total scalar potential can experience a diverse and complex array of phase transition processes with the thermal corrections. The FOEWPT requires the existence of two degenerate vacua, one with electroweak symmetric $\langle h\rangle=0$, the other with the broken one $\langle h \rangle \neq0$ at some critical temperature $T_c$. In the early universe, with extremely high temperature $T\gg T_c$, the universe falls into the symmetric phase. Then, as the universe expands and cools, when the temperature drops below $T_c$, the symmetric vacuum $\langle h\rangle=0$ is no longer the global minimum; instead, the broken vacuum $\langle h\rangle\neq 0$ becomes the much deeper minimum, and the universe will gradually decay to the broken phase $\langle h\rangle\neq 0$.
The decay rate per unit volume is \cite{Linde:1981zj}
\begin{equation}
\Gamma(T)\sim T^4\left(\frac{S_3(T)}{2\pi T}\right)^{3/2}e^{-S_3(T)/T},
\end{equation}
where $S_3(T)$ represents the classical action of the $O(3)$-symmetric bounce solution. A phase transition takes place when the decay rate per Hubble volume reaches the $O(1)$. The nucleation temperature $T_n$ is determined using the cosmological Kibble-Zurek criterion: $\Gamma(T_n)=H^4(T_n)$, where $H(T)$ is the Hubble constant at temperature $T$.
In a Universe dominated by radiation, where a FOPT occurs near the EW scale, the temperature $T_n$ can be determined using the following approximate relation \cite{Quiros:1999jp}:
\begin{equation}
S_3(T_n)/T_n\approx140,
\end{equation}
which serves as the criterion for an FOEWPT. For each parameter input in Eq. (\ref{eq:param}), after substituting into Eq. (\ref{eq:coupling-Tem}) and Eq. (\ref{eq:corr-potent}), the nucleation temperature $T_n$ can be calculated using the {\tt Python} package {\tt cosmoTransitions} \cite{Wainwright:2011kj}.

Stochastic GWs are generated by FOEWPT mainly through three processes: bubble collisions \cite{Kosowsky:1991ua,Kosowsky:1992vn,Huber:2008hg,Bodeker:2009qy,Jinno:2017ixd}, sound waves in the plasma \cite{Hindmarsh:2013xza,Hindmarsh:2015qta}, and magnetohydrodynamics turbulence \cite{Hindmarsh:2013xza,Hindmarsh:2015qta}.
The present-day GW spectrum can be calculated 
as the spectrum sum of the three sources, that is $\Omega_{\rm GW}h^2=\Omega_{\rm bc}h^2+\Omega_{\rm sw}h^2+\Omega_{\rm turb}h^2$, which can be well respectively expressed as functions of a series of parameters, such as the energy budget of FOEWPT normalized by the radiative energy $\alpha$, the inverse time duration of the FOEWPT $\beta$, the bubble wall velocity $v_w$, and other relevant parameters \cite{Huber:2008hg,Hindmarsh:2015qta,Espinosa:2010hh,Caprini:2009yp,Binetruy:2012ze} (see Ref. \cite{Shibuya:2022xkj} for a summary of the specific computational details). Among these parameters, two crucial parameters for the GW are the energy budget of FOEWPT normalized by the radiative energy $\alpha$ and the inverse time duration of the FOEWPT $\beta$, which are given by
\begin{equation}
\alpha=\frac{1}{\rho_R}\left(T\frac{\partial\Delta V_T}{\partial T}-\Delta V_T\right)\Big|_{T_n};\quad \beta/H=T_n\frac{d(S_{3}/T)}{dT}\Big|_{T_n},
\end{equation}
where $\rho_R=g_*\pi^2T_n^4/30$ is the radiation energy of the thermal bath, $\Delta V_T$ is the effective potential difference between the true and false vacua, and $g_*\sim100$ is the number of relativistic degrees of freedom.
\begin{figure}[ht]
    \centering
    \includegraphics[width=0.49\textwidth]{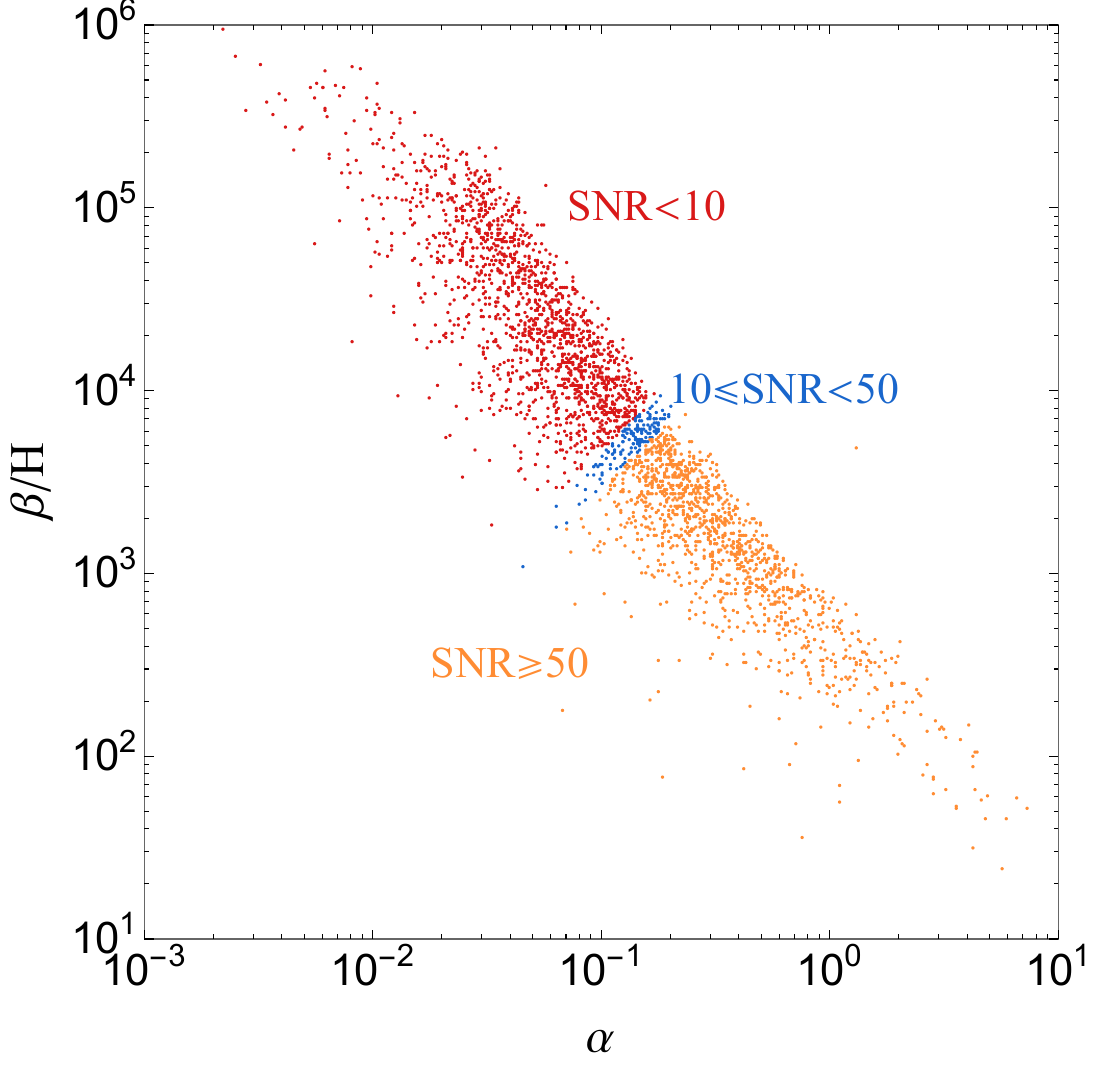}
    \includegraphics[width=0.49\textwidth]{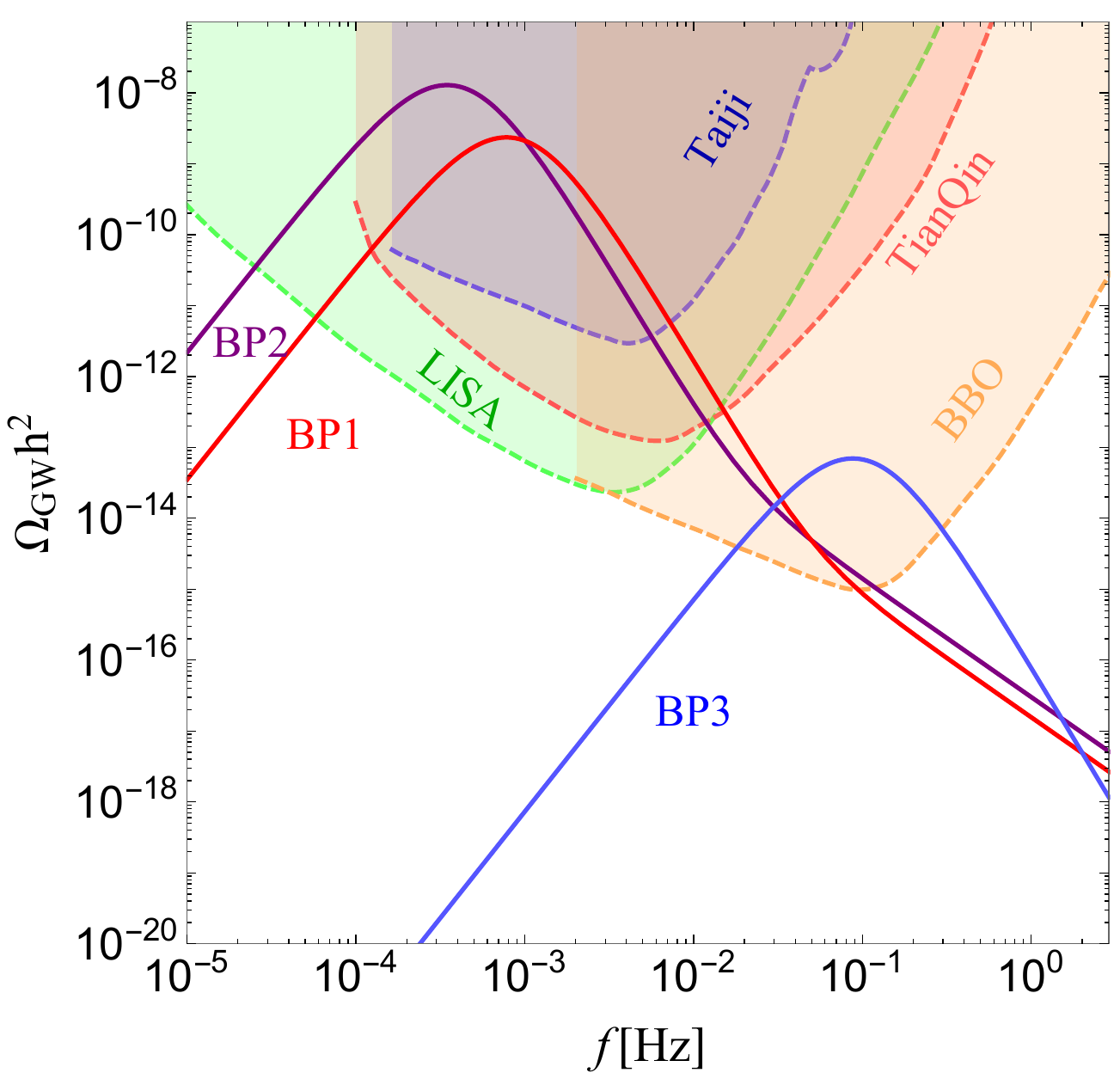}
    \caption{{\bf Left}: The scanned results in $\alpha-\beta/H$ plane for three different SNR regions, where red dots refer to ${\rm SNR}<10$, blue ones represent $10\leqslant{\rm SNR}<50$, and orange ones characterize ${\rm SNR}\geqslant50$. {\bf Right}: The gravitational wave energy abundance, $\Omega_{\rm GW}h^2$, as a function of the GW frequency, where the green, red, blue, and orange shaded regions represent the projected sensitivity of LISA, TianQin, Taiji, and BBO, respectively. The blue, red and purple solid lines refer to the possible GW spectrum from three benchmark points.}
    \label{fig:SNR-alpha-beta}
\end{figure}

Once the total GW spectrum for each parameter input is calculated, the signal-to-noise ratio (SNR) \cite{Caprini:2015zlo} provides a common method to estimate the detectability of the GW signal at an interferometer. Several interferometers, such as LISA \cite{Klein:2015hvg}, BBO/DECIGO \cite{Kudoh:2005as,Yagi:2011wg}, TianQin \cite{TianQin:2015yph}, and Taiji \cite{Gong:2014mca} programs, serve as excellent platforms for GW observation. Taking LISA as an example, the SNR can be expressed as \cite{Caprini:2015zlo}
\begin{equation}
{\rm SNR}=\sqrt{t\int_{f_{\rm min}}^{f_{\rm max}} df\left(\frac{\Omega_{\rm GW}(f)}{\Omega_{\rm LISA}(f)}\right)^2},
\end{equation}
where $\Omega_{\rm LISA}$ represents the power spectrum density of the LISA detector \cite{Caprini:2015zlo}, and $t=9.46 \times 10^7~{\rm s}$ signifies a typical operational duration (approximately four years) \cite{Caprini:2019egz}. 

In the left panel of Fig. \ref{fig:SNR-alpha-beta}, we illustrate the impact of various $\alpha$ and $\beta/H$ distributions on the SNR results. We find that the SNR progressively increases with higher $\alpha$ values while it decreases with lower $\beta/H$ values. 
This can be attributed to the necessity for a more rapid and vigorous phase transition in the presence of a stronger FOPT.
To achieve a sufficiently large SNR that GW detectors can detect, the parameter space is predominantly concentrated in the region where $\alpha > 0.1$ and $\beta/H < 10^4$.
And the predicted GW spectra, $\Omega_{\rm GW}h^2$, of the FOEWPT, as a function of frequency $f$, are shown in the right panel of Fig. \ref{fig:SNR-alpha-beta} for three benchmark points (BPs):
\begin{equation}
\begin{aligned}
    {\rm BP1}: ~ &m_s = 600~ {\rm GeV},~ \theta_{hs} = 0.1, ~ v_s = 700~{\rm GeV},~ \lambda_{HS} = 7.9, ~ \lambda_s = 0.5, \\
    &\text{which gives~}T_n = 37.5~{\rm GeV},~ \alpha = 1.1, ~ \beta/H = 102.7,\\
    {\rm BP2}: ~ &m_s = 760~ {\rm GeV},~ \theta_{hs} = 0.1, ~ v_s = 500~{\rm GeV},~ \lambda_{HS} = 12, ~ \lambda_s = 1.5, \\
    &\text{which gives~}T_n = 30~{\rm GeV},~ \alpha = 2.6, ~ \beta/H = 59,\\
    {\rm BP3}:~&m_s = 600 ~{\rm GeV},~ \theta_{hs} = 0.1, ~ v_s = 29~{\rm GeV},~ \lambda_{HS} = 5.8, ~ \lambda_s = 1.5, \\
    &\text{which gives~}T_n = 85.5 ~{\rm GeV}, ~ \alpha = 0.1, ~ \beta/H = 5\times10^{3}.
\end{aligned}
\end{equation}
Besides, the sensitivity curves of several GW experiments, such as LISA, TianQin, Taiji, and BBO, are also plotted in this figure.
It can be seen that, the GW signal from BP3 is roughly three orders of magnitude weaker than that from BP1 and BP2, primarily because $\alpha_{\rm BP1,BP2}>\alpha_{\rm BP3}$ and $\beta/H_{\rm BP1,BP2}<\beta/H_{\rm BP3}$.
Returning to the physical parameters, for sufficiently large values of $v_s$ and $m_s$, the resulting GW strength surpasses the sensitivity curves of detectors over large frequency ranges, producing potentially observable signals. By contrast, small $v_s$ and $m_s$ substantially suppress the spectrum so that only a portion lies within BBO’s sensitivity, which makes detection challenging. Therefore, in the parameter space we focus on ($m_N > 100 ~{\rm GeV}$, which means $v_s\gtrsim m_N\sim100$ GeV), current and next-generation GW detectors can provide significant constraints on the model, and improved detector sensitivity would further tighten these limits.

\section{Sterile Neutrino Searches at the HL-LHC}
\label{sec:sensi}
As demonstrated in the last section, the frequency and strength of the GWs from the FOPT are strongly relevant to the mixing strength $\lambda_{HS}$, which can result in a moderate interaction strength between the $s$ and SM particles after rotation to the mass eigenstates, as derived in Eq. (\ref{eq:s-h-coupling}). 
This interaction enables the production and detection of $s$ particles at colliders. 
The interactions of sterile neutrino to active neutrino and $s$ make probing sterile neutrino at colliders promising, where the sterile neutrino can be substantially produced through $s$ particle decays. Due to suppression by the tiny mixing angle $\theta_{\nu N}$, the sterile neutrino will subsequently decay with a long lifetime, leaving a displaced signature at the colliders. This signature will be explored in this section.

Although it has been discussed at the beginning of Sec. \ref{sec:constraints} that the sterile neutrino-related parts are subject to almost no constraints due to the tiny mixing angle with the active neutrino, we can temporarily set aside the explanation of the active neutrino mass issue. This allows a model-independent exploration of sterile neutrinos at colliders. 
In the model-independent perspective, the mixing element $V_{\ell N}$ (same as $\theta_{\nu N}$) and the sterile neutrino mass $m_N$ are generally treated as independent parameters. In Refs. \cite{CMS:2024xdq,CMS:2022hvh}, the authors investigated and summarized the experimental limits on $V_{\ell N}$ as a function of the sterile neutrino mass $m_N$, which usually applied its prompt decay searches or vector boson fusion via a $t$-channel off-shell heavy sterile neutrino. These experiments place soft restrictions on 
$V_{\ell N}$ for the heavy sterile neutrino sector; for example, $|V_{\ell N}|^2<10^{-3}$ is required when $m_N>100$ GeV.

In our model, since we are interested in sterile neutrinos with mass $m_N\gtrsim 100$ GeV, the LHC serves as a natural search platform. In particular, the relatively weak experimental constraints on the $s$-$h$ mixing angle allow for a substantial coupling strength between $s$ and gluons, thereby facilitating the production of a pair of sterile neutrinos via the $s$-channel process mediated by the $s$ boson in gluon fusion. Since the $N\to W^\pm \ell^\mp$ channel has the largest branching ratio, we will try to investigate its collider signatures. In the following discussions, we take the mixing between $N$ and the second-generation lepton $\nu_\mu$ as an example. An inclusive searching strategy is applied, which requires at least one sterile neutrino to decay within the detector volume, leaving a displaced vertex.
The corresponding full signal process is:
\begin{align}\label{eq:process}
    p p \xrightarrow{s^{(*)}/h^*} (j)~ N N, ~N\to W^\pm \mu^\mp, ~W^\pm \to jj,
\end{align}
where $j$ with a parenthesis means two parton-level processes, with or without an initial radiation jet. 
The number of signal events at the HL-LHC with a center-of-mass energy $\sqrt{s}=14$ TeV and an integrated luminosity $\mathcal{L}=3~{\rm ab^{-1}}$ can be written as
\begin{equation}
    N=\mathcal {L}\cdot\sigma\cdot \langle \mathbb{P}\cdot \epsilon \rangle,
\end{equation}
where $\sigma$ is the cross section for processes of Eq. (\ref{eq:process}), $\mathbb{P}$ is the probability for one $N$ decaying within a given distance and $\epsilon$ is the kinematical cut efficiency. $\langle \mathbb{P}\cdot \epsilon \rangle$ is the averaged inclusive searching efficiency, which is calculated event-by-event.

To numerically derive the cut efficiency, a dedicated signal event Monte Carlo simulation is applied, which can be divided into the following steps. Firstly, parton-level events are generated using {\tt MadGraph 5} \cite{Alwall:2014hca}, by entering the UFO model generated by {\tt FeynRules} \cite{Alloul:2013bka} based on the whole UV model in Sec. \ref{sec:model}. Secondly, the events are passed to the {\tt Pythia8} \cite{Sjostrand:2007gs} and {\tt Delphes} \cite{deFavereau:2013fsa} for showering, hadronization, and detector effects by use of the standard CMS card.
The jets are clustered using the anti-$k_t$ and fat jet \cite{Heinrich:2014kza} algorithm with radius parameter $R$ = 0.5 ($R$ = 0.8 for fat jets).

For a given event, the probability that a long-lived sterile neutrino, $N$, decays within a specific range $[r_1\cdot {\bf \hat r},~r_2\cdot {\bf \hat r}]$ along its moving direction ${\bf \hat r}$ can be derived as \cite{Cao:2023smj}:
\begin{equation}
    \mathbb{P}=\exp \left(-\frac{r_1}{\gamma\beta c \tau_N} \right)-\exp\left(-\frac{r_2}{\gamma\beta c \tau_N}\right),
\end{equation}
where $\gamma$ is the Lorentz factor of sterile neutrino $N$, $\beta$ is its speed, and $\tau_N$ is its proper lifetime. With this, one can calculate the probability that the displaced distance $d_N$ of a sterile neutrino satisfies $0.4 ~{\rm cm}<|d_N\cdot \sin \alpha'|< 30 ~{\rm cm}$ \cite{ATLAS:2017tny}, where $\alpha'$ is the angle between the sterile neutrino moving direction and the beamline axis. The lower bound can effectively reduce the SM backgrounds from prompt decays, and the upper bound is required to ensure the good reconstruction efficiency of SM charged particles from sterile neutrino decays. In addition, to further eliminate the SM backgrounds, the invariant mass of reconstructed displaced jets is required to be close to the $W$ mass. We can also require the invariant mass of all displaced final charged particles to be around the sterile neutrino mass. With additional specific kinematic requirements, the complete selection criteria can be organized as \cite{ATLAS:2017tny,Chiang:2019ajm}:
\begin{equation}\label{eq:DV-cuts}
\begin{aligned}
    {\rm DV}:~~
    &p_T^{\mu}>20 ~{\rm GeV},~|\eta_{\mu}|<2.5,~p_T^{j,~{\rm trigger}}>120 ~{\rm GeV},~p_T^j>20 ~{\rm GeV},~|\eta_j|<2.5,\\
    &0.4~{\rm cm}<|d_N\cdot \sin\alpha'|<30~{\rm cm},~|d_N\cdot \cos\alpha'|<30~{\rm cm},\\
    & m_{jj}~(m_{j}^{\rm fat})\in [50,~100]~{\rm GeV}, ~m_{\mu jj} ~(m_{\mu j^{\rm fat}})\in [0.8m_N,~1.2m_N],
\end{aligned}
\end{equation}
where the particles appeared are required to be well recognized and reconstructed in the Delphes, $p_T^{\mu/j}$ and $\eta_{\mu/j}$ are the transverse momentum and pseudorapidity of the muon/jets from $N/W$ decays, respectively. $p_T^{j,~{\rm trigger}}$ is the transverse momentum of the hardest (initial) jet, which is used to trigger the event.  
And $m_{jj}$ ($m_{j}^{\rm fat}$) is the invariant mass of the jets (fat jet) from $W$ boson decays. $m_{\mu jj}$ or $m_{\mu j^{\rm fat}}$ is the invariant mass of all the particles from $N$ decays.
Since our simulation is performed at the detector level, the isolation and identification of muons and jets are automatically considered.

These selection criteria can not only ensure the signal reconstruction, but also suppress the SM backgrounds to a negligible level. As discussed in Refs. \cite{Liu:2020vur,Bi:2024pkk}, the energetic well-reconstructed displaced muon tagging and narrow invariant mass cuts on $W$ boson and sterile neutrino $N$ effectively remove the remaining SM backgrounds from the interaction of SM particles with the detector materials, QCD hadron and meson events, tau decays, and fake-track coincidental background. A dedicated analysis of these backgrounds is beyond the scope of this work and will be discussed in more detail in future studies.

\begin{figure}[ht]
    \centering
    \includegraphics[width=0.49\textwidth]{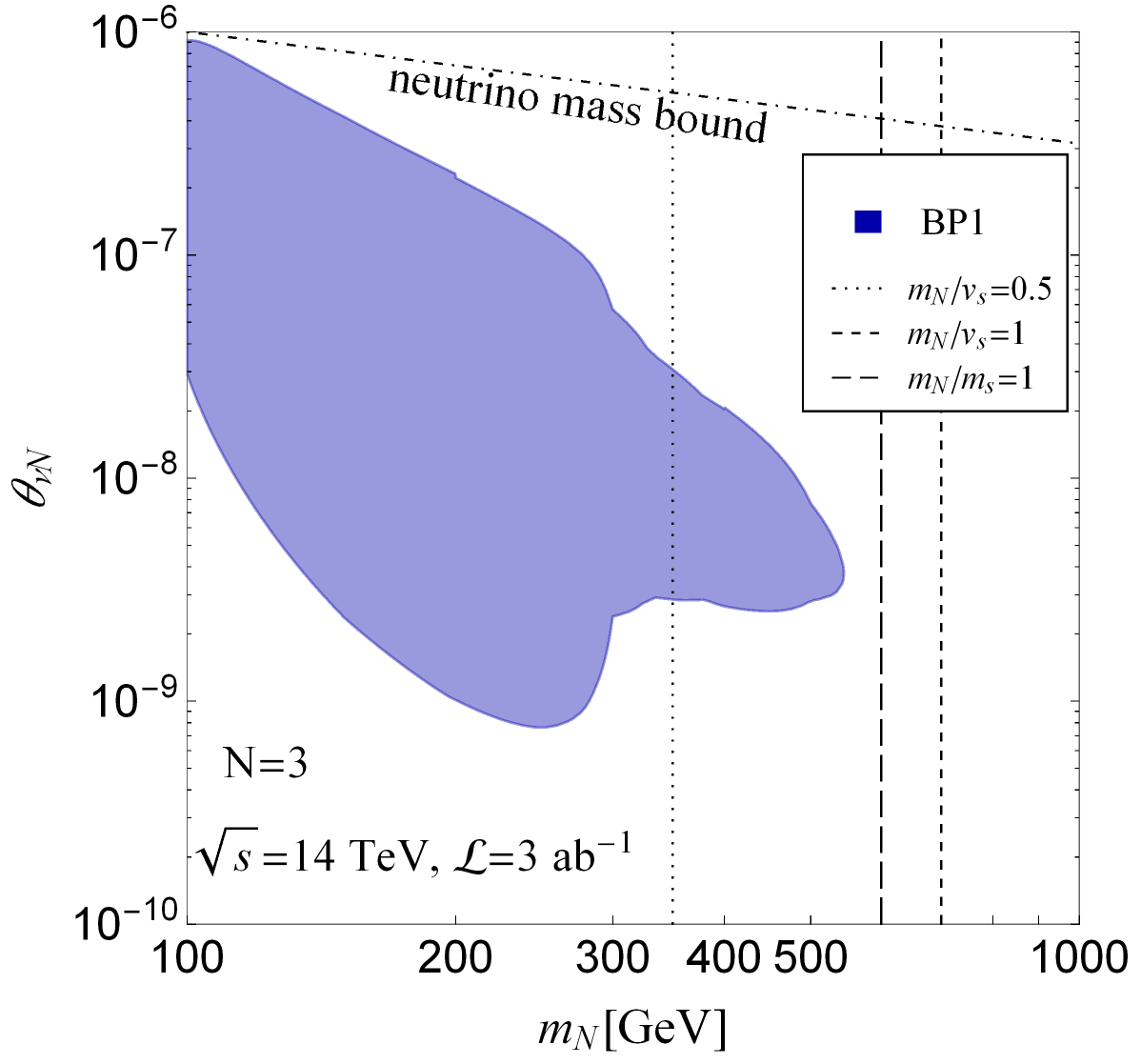}
    \includegraphics[width=0.49\textwidth]{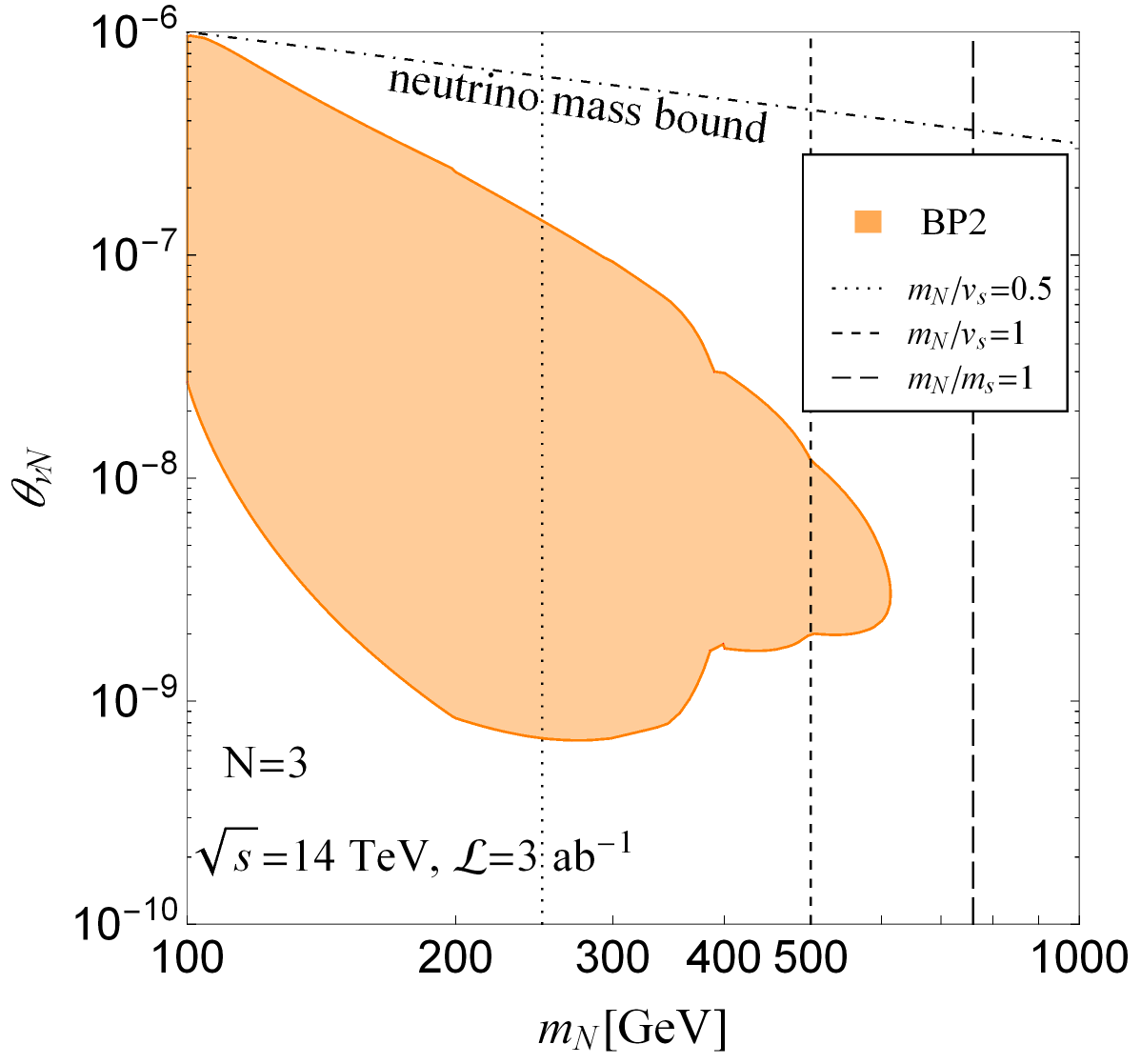}
    \caption{The projected 95\% C.L. sensitivities for inclusive displaced vertex searches for $pp\to (j)NN$ with sterile neutrino further decay to $\mu W$ as a function of sterile neutrino mass $m_N$ at HL-LHC are shown in the color-shaded regions with the center-of-mass energy $\sqrt{s}=14$ TeV and integrated luminosity $\mathcal{L}=3~{\rm ab^{-1}}$.
    {\bf Left}: The benchmark BP1: $\{\theta_{hs}=0.1,~m_s=600~{\rm GeV}, ~{v_s=700~{\rm GeV}},~\lambda_{HS} = 7.9, ~ \lambda_s = 0.5\}$.
    {\bf Right}: The benchmark BP2: $\{\theta_{hs}=0.1,~m_s=760~{\rm GeV}, ~{v_s=500~{\rm GeV}},~\lambda_{HS} = 12, ~ \lambda_s = 1.5\}$.
    Besides, several dotted and dashed lines are also drawn, where the three vertical lines correspond to three typical values, $y_N\simeq\frac{m_N}{v_s}$ equals 0.5, 1, and $\frac{m_N}{m_s} =1$. The other oblique line spanning from left to right represents the upper limit of the active neutrino mass requirement. }
    \label{fig:sensi}
\end{figure}

The 95\% C.L. sensitivities for probing the long-lived sterile neutrino, $N$, at 14 TeV HL-LHC with an integrated luminosity $\mathcal{L}=3~{\rm ab}^{-1}$ are shown in Fig.~\ref{fig:sensi}, where the 95\% C.L. corresponds to 3 signal events under the zero SM backgrounds assumption. 
In the left panel of Fig~\ref{fig:sensi}, the benchmark is considered with parameters set as $\{\theta_{hs}=0.1,~m_s=600~{\rm GeV}, ~{v_s=700~{\rm GeV}}, ~\lambda_{HS} = 7.9, ~ \lambda_s = 0.5\}$.
It is observed that when the mass of the sterile neutrino $m_N<m_s/2$ with $m_s$ fixed to 600 GeV, the collider demonstrates enhanced sensitivity compared to scenarios where $m_N$ exceeds $m_s/2$. This increased sensitivity is attributed to the larger production cross section for sterile-neutrino pairs, since in this $m_N<m_s/2$ mass range the pairs can be produced via decays of the on-shell scalar $s$. 
As $m_N$ increases beyond $m_s/2$, production must proceed via off-shell $s$, which leads to the decline in sensitivity. 
This kinematic transition produces a pronounced inflection in the sensitivity curves around $m_N \sim m_s/2$, corresponding to $m_N = 300 ~{\rm GeV}$.
Moreover, there is a significant decrease in sensitivity within the region $m_N\lesssim 200$ GeV owing to diminished cut efficiency. 
More precisely, a light sterile neutrino is particularly boosted, and so are its decay products, which leads to collimated jets that are more difficult to reconstruct after being processed through Pythia8 and Delphes.
Especially, for $m_N$ around 100 GeV, the efficiency is severely reduced, because reconstructing the muon from highly boosted $N$ becomes challenging due to the difficulty in isolating muons from jets according to our simulations. 
The dot–dashed line denotes the boundary of parameter space that yields neutrino masses compatible with existing experimental constraints; the region above this line is excluded by neutrino-mass measurements.
The vertical lines indicate $m_N/v_s$ corresponding to the Yukawa coupling between $N$ and $s$, and $m_N/m_s$ corresponding to the opening of $N \rightarrow s \nu$ decay channel. All the sensitivity region satisfies $m_N/v_s<1$ and $m_N<m_s$, so that we can neglect the decay process of $N \rightarrow s \nu$.
For this benchmark, the HL-LHC can probe mixing angles $\theta_{\nu N}$ between active neutrino and sterile neutrino within $[6\times10^{-10},~10^{-6}]$ and the sterile neutrino mass $m_N$ in the range $[100,~500]$ GeV.

In the right panel of Fig. \ref{fig:sensi}, the other benchmark scenario is examined with the parameters specified as $\{\theta_{hs}=0.1,~m_s=760~{\rm GeV},~v_s=500~{\rm GeV},~\lambda_{HS} = 12,~\lambda_s = 1.5\}$. At the HL-LHC, the sensitivity for detecting $\theta_{\nu N}$ can reach as low as $6\times10^{-10}$, provided the sterile neutrino mass is below $380~{\rm GeV}$. Above the region where $m_N \sim m_s/2$, the sensitivity drops considerably due to a decrease in the signal production cross section, similar to the behavior observed in BP1. For this benchmark, the sensitivity to sterile neutrino masses spans the range $m_N \in [100,~600]$ GeV, with corresponding mixing angles $\theta_{\nu N} \in [6\times10^{-10},~10^{-6}]$.

\section{The interplay between GW and collider searches}
\label{sec:interplay}
Our model for the sterile neutrino and extended Higgs sector involves 7 free parameters, as delineated in Eq. (\ref{eq:param}). Our investigation reveals that although GW detections and collider experiments display diverse levels of sensitivity to these parameters, they can be fruitfully integrated for conjoint analyses when specific parameters are fixed. This synergy creates an opportunity to employ both GW observations and collider searches as complementary methods to corroborate and affirm the validity of the model. 
\begin{figure}[ht]
    \centering
    \includegraphics[width=0.48\linewidth]{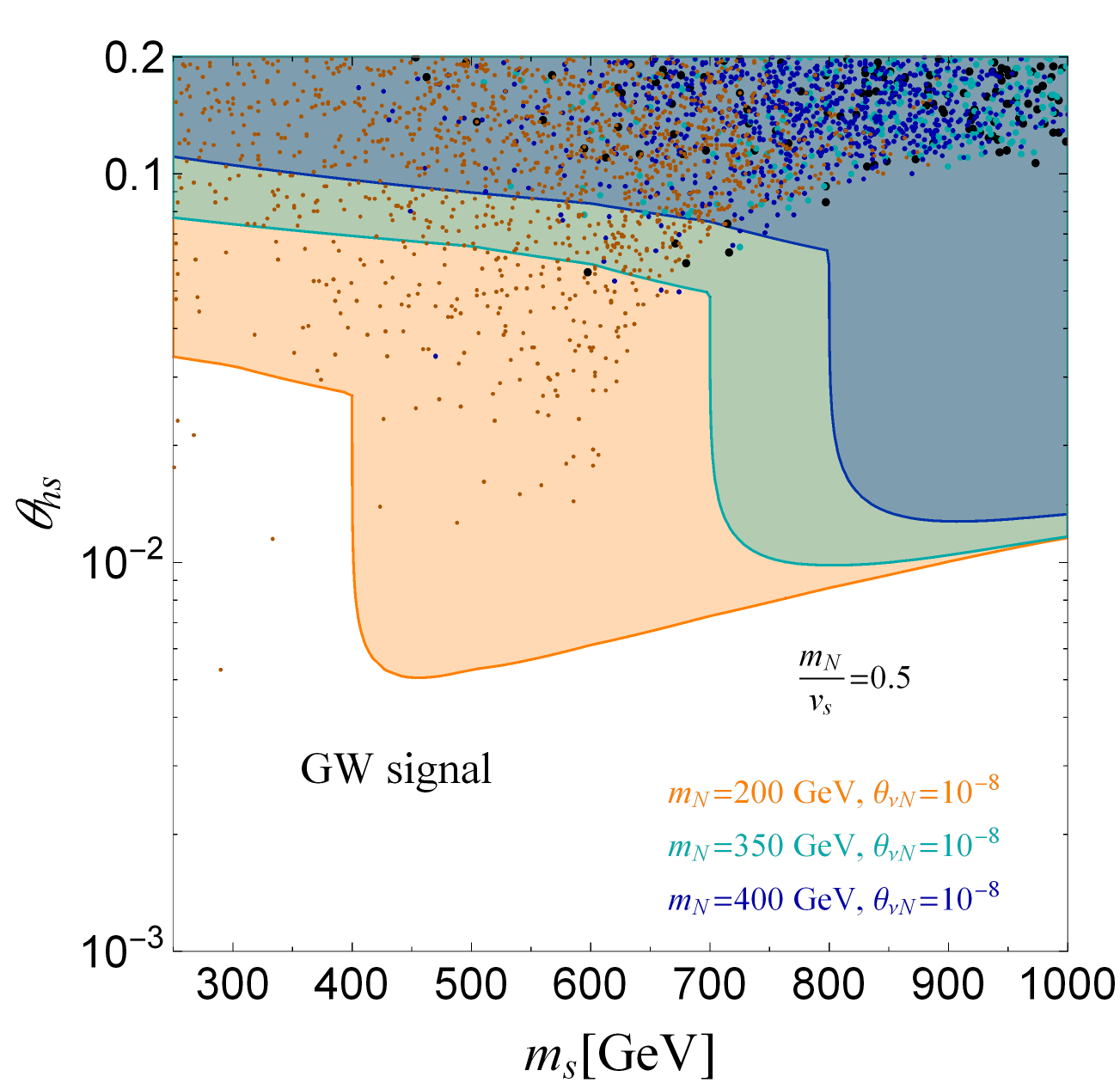} ~~~~
    \includegraphics[width=0.48\linewidth]{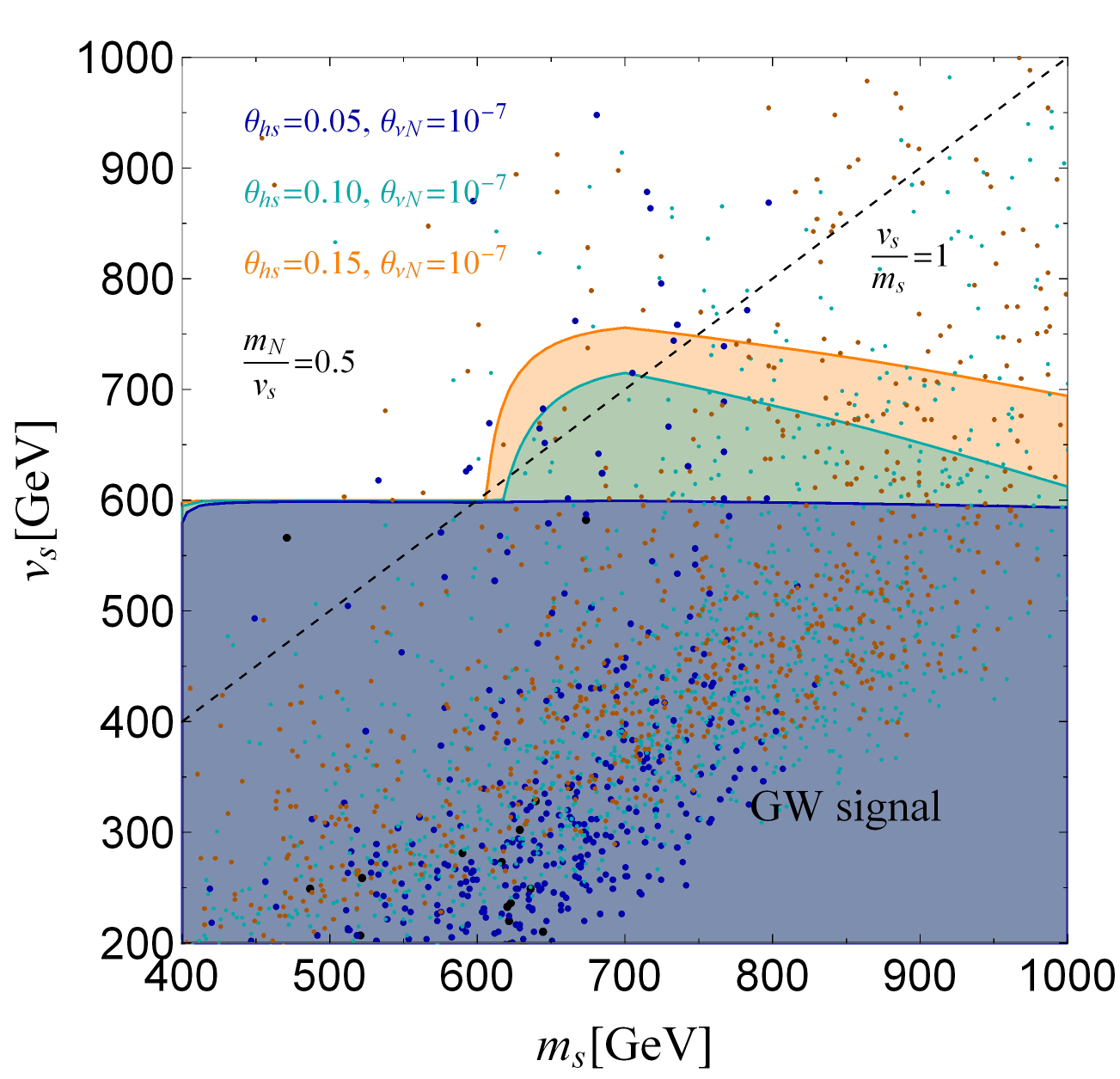}
    \caption{The interplay between gravitational wave detection and LHC searches. The dots correspond to the LISA-detectable gravitational wave signals, and the color-shaded regions are the HL-LHC projections at $\sqrt{s}=14$ TeV with $\mathcal{L}=3~{\rm ab^{-1}}$. 
    {\bf Left}: The projected joint searches for the parameters, $\theta_{hs}-m_s$, with three sterile neutrino masses fixed. The points are the GW data, in which dark orange, cyan, blue, and black colors represent $v_s \in [200,400)~{\rm GeV},~ [400,700)~{\rm GeV},~[700,800)~{\rm GeV}$ and $[800,1000)$ GeV.
    {\bf Right}: The projected searches for the parameters, $v_s-m_s$, with three Higgs-scalar mixing angles fixed. The points are the GW data, in which black, dark blue, cyan, and orange colors represent $\theta_{hs} \in [0,0.05),~ [0.05,0.1),~[0.1,0.15)$ and $[0.15, 0.2)$.} 
    \label{fig:GW-collider}
\end{figure}

In Fig. \ref{fig:GW-collider}, we present the GW signal data points together with the collider search regions, highlighting correlations among the parameter spaces of $m_s, v_s$, and $\theta_{h s}$.
The 95\% C.L. sensitivities of HL-LHC searches to long-lived sterile neutrinos are presented, and GW observations by LISA using a detection threshold of SNR > 10. This juxtaposition underscores the combined power of gravitational-wave and collider studies in probing and placing constraints on the sterile neutrino sector.

The left panel of Fig. \ref{fig:GW-collider} illustrates the relationship between GW observations and HL-LHC searches within the $m_s-\theta_{hs}$ plane. In our collider investigation, we choose benchmark points of sterile neutrino masses of $m_N = 200, 350, 400~\mathrm{GeV}$, paired with a mixing angle of $\theta_{\nu N} = 10^{-8}$. Without loss of generality, we choose the ${\rm BR}(s\rightarrow NN)\approx 0.5$. The HL-LHC sensitivity is depicted through various color-shaded regions. As $m_N$ is linked to the Yukawa coupling $y_N$ and $v_s$, the condition $m_N / v_s = 0.5$ sets a specific value for $v_s$ for each $m_N$ considered.
For fixed $m_N$, from left to right, as $m_S$ increases, the production cross section of $N$ pair mediated by off-shell $s$ gradually grows. Consequently, the colored region expands and extends into progressively smaller $\theta_{hs}$. It keeps growing until it reaches around $m_s \sim 2m_N$, at which point an abrupt enhancement occurs, because $N$ can then be produced through the decay of an on-shell $s$. Alternatively, for a fixed $m_s$, a smaller $m_N$ yields a larger production cross section, thereby increasing sensitivity and enabling probes of smaller $\theta_{hs}$ regions of the parameter space. In parallel, the projected sensitivities of future GW observatories are shown as colored bands corresponding to different benchmark scenarios.
The points have been assigned distinct colors based on $v_s$, with black indicating $800~{\rm GeV} \leqslant v_s < 1000~{\rm GeV}$, dark blue denoting $700~{\rm GeV} \leqslant v_s < 800~{\rm GeV}$, dark cyan representing $400~{\rm GeV} \leqslant v_s < 700~{\rm GeV}$, and dark orange signifying $200~{\rm GeV} \leqslant v_s < 400~{\rm GeV}$.

In the right panel of Fig. \ref{fig:GW-collider}, analogous patterns are employed to illustrate the relationship between GW signals and HL-LHC projections within a different parameter space, $v_s-m_s$. Black data points signify cases where $0 \leqslant \theta_{hs} < 0.05$, while dark orange points represent $\theta_{hs} \geqslant 0.15$. Dark cyan points correspond to the range $0.1 \leqslant \theta_{hs} < 0.15$, and dark blue points to $0.05 \leqslant \theta_{hs} < 0.1$. The shaded regions in color indicate the HL-LHC sensitivity. We examine three scenarios with $\theta_{hs} = 0.05,~0.1,~0.15$ and a fixed $\theta_{\nu N} = 10^{-7}$. The dashed line signifies $v_s/m_s = 1$, which marks the boundary between on-shell and off-shell $s$-mediated $N$ pair production when $m_N/v_s = 0.5$.
In light $m_s$ range, the sensitivity boundary is reached at $v_s = 600~{\rm GeV}$, corresponding to $m_N = 300~{\rm GeV}$. This is because the particle $N$ generates a displaced-vertex signature predominantly at this mass, with $\theta_{\nu N}$ held constant at $10^{-7}$. As $m_N$ ($v_s$) increases, the decay width broadens, leading to prompt decays. For $m_s > 600~{\rm GeV}$, the production cross section is augmented due to on-shell $s$ production, thereby extending the explored region of the parameter space. Moreover, the range of accessible $v_s$ values increases with a higher $\theta_{hs}$. Consequently, within this domain, raising $\theta_{hs}$ results in a discernible protrusion in the sensitivity map.

Collider searches face constraints when exploring the parameter space at small values of $\theta_{h s}$, as the production cross-section is significantly reduced in this regime. In scenarios where $m_N/v_s$ and $\theta_{\nu N}$ are held constant, the HL-LHC can only probe up to $v_s = m_N / 0.5 = 800~ \mathrm{GeV}$ when $m_N$ is 400 GeV. At scales beyond this, collider sensitivity decreases dramatically. Conversely, GW signals, represented by points, remain detectable in areas where collider searches become ineffective, especially for higher scalar vacuum expectation values.

The regions where GW signal points coincide with collider sensitivity bands, indicated by matching colors, delineate a parameter space that can be investigated by both experimental approaches concurrently. This overlap facilitates strong cross-validation of this model. On the other hand, areas without overlap pinpoint sectors that are exclusively accessible to either GW or collider investigations. Collectively, these findings indicate that within specific feasible parameter domains, GW and collider experiments offer a synergistic exploration of the sterile neutrino parameter space, establishing an effective methodology for model verification.

\section{Conclusions}
\label{sec:conclusion}

In this work, a UV-complete model with a sterile neutrino obtaining mass via the spontaneous symmetry breaking of a new scalar is explored, where the new introduced scalar, combined with the electroweak Higgs field, can not only generate observable GWs through the FOEWPT, but also provide a promising sterile neutrino production channel at HL-LHC via the scalar mixing. Furthermore, this model addresses the neutrino mass problem following the traditional see-saw mechanism, which also naturally yields the sterile neutrino decay channel. The suppression of this decay by the tiny mixing angle between the active and sterile neutrinos leads to a long-lived signature at the collider length scales.

We first investigated the phase transition of the combined potential $V(\hat H, \hat S)$ of the new scalar and Higgs field in the early universe. Specifically, we derived the thermal corrected effective potential and embedded it into the cosmoTransitions to pinpoint the available regions for parameter spaces that support a detectable first-order electroweak phase transition. Next, we obtained the feasible FOEWPT for different parameter spaces $m_s$ and $v_s$ after fixing the scalar mixing angle $\theta_{hs}$, which are densely populated in the area with small $v_s$. Via simulations, the GW energy spectrum $\Omega_{\rm GW}h^2$ associated with characteristic parameters $\alpha$ and $\beta$ is obtained, showing that the predicted frequency and strength lie within the sensitivity of LISA, BBO/DECIGO , Taiji, and TianQin observations.  

Secondly, we explore the long-lived signatures of the sterile neutrino $N$ at the HL-LHC, especially focusing on inclusive searches for the generation of pair sterile neutrinos via the $s$-channel mediated by the Higgs and the new scalar $s$. The on-shell production of heavy scalar $s$ and its decay into the sterile neutrino pair greatly enhances the production rate for $N$. Due to the larger decay branching ratio, the $N\to W\mu$ channel is adopted in the analysis. Utilizing the displaced vertex method, we explore the detection of the long-lived sterile neutrino. We find that the 14 TeV HL-LHC with an integrated luminosity $\mathcal{L}=3{~\rm ab^{-1}}$ exhibits significant sensitivities to sterile neutrino mass in the range $m_N\in [100,~500]$ GeV, with mixing angle $\theta_{\nu N}\in[6\times10^{-10},10^{-6}]$ for the benchmark point $\{\theta_{hs}=0.1,~m_s=600~{\rm GeV}, ~{v_s=700~{\rm GeV}},~\lambda_{HS} = 7.9, ~ \lambda_s = 0.5\}$ and $m_N\in [100,~600]$ GeV with mixing angle $\theta_{\nu N}\in[6\times10^{-10},10^{-6}]$ for the benchmark point $\{\theta_{hs}=0.1,~m_s=760~{\rm GeV}, ~{v_s=500~{\rm GeV}},~\lambda_{HS} = 12, ~ \lambda_s = 1.5\}$. The sensitivities vary slightly with different benchmark points, but the overall sensitive regions remain largely unchanged. Additionally, combined sensitivities from the GW observations and collider searches are also investigated.
We find that most of the parameter space can be cross-checked by both GW and HL-LHC searches, while others can only be exclusively detected through either gravitational wave observations or collider searches. Consequently, LHC searches and GW detection exhibit strong complementarity.

\section{Acknowledgments}
We would like to thank Ke-Pan Xie for useful discussions and sharing the codes.
The work of J.G is supported by the Postdoctoral Fellowship Program (Grade C) of China Postdoctoral Science Foundation under Grant No. GZC20252775.
The work of J.L. is supported by the National Science Foundation of China under Grant No. 12235001 and No. 12475103.
The work of X.P.W. is supported by National Science Foundation of China under Grant No. 12375095, and the Fundamental Research Funds for the Central Universities.
J.L. and X.P.W. thank APCTP, Pohang, Korea, for their hospitality during the focus program [APCTP-2025-F01], from which this work greatly benefited. J.L. and X.P.W. also
thank the Mainz Institute for Theoretical Physics (MITP) of the PRISMA+ Cluster of Excellence
(Project ID 390831469) for its hospitality and partial support during the completion of this work. The authors gratefully acknowledge the valuable discussions and insights provided by the members of the Collaboration of Precision Testing and New Physics.

\bibliographystyle{JHEP}
\bibliography{ref}
\end{document}